\documentclass[showpacs,preprintnumbers,amsmath,amssymb,prl,superscriptaddress,twocolumn,longbibliography,footinbib]{revtex4-1}

\usepackage{natbib}
\usepackage{txfonts}
\usepackage{graphicx}
\usepackage{dcolumn}
\usepackage{bm}
\usepackage{hyperref}
\usepackage{color}

\usepackage{amssymb}
\usepackage{amsmath}
\usepackage{float}
\usepackage{epsfig}
\usepackage{blindtext}

\newcommand{\ket}[1]{| #1 \rangle}
\newcommand{\bra}[1]{\langle #1 |}

\usepackage[utf8]{inputenc}

\usepackage{silence}
\newcommand{\kn}{{n\mathbf{k}}}
\begin{document}
\title{Strongly enhanced Berry dipole at topological phase transitions in BiTeI} 

\author{Jorge I. Facio}
\affiliation{Institute for Theoretical Solid State Physics, IFW Dresden, Helmholtzstr.~20, 01069 Dresden,
  Germany}

\author{Dmitri Efremov}
\affiliation{Institute for Theoretical Solid State Physics, IFW Dresden, Helmholtzstr.~20, 01069 Dresden,
  Germany}

\author{Klaus Koepernik}
\affiliation{Institute for Theoretical Solid State Physics, IFW Dresden, Helmholtzstr.~20, 01069 Dresden,
  Germany}

\author{Jhih-Shih You}
\affiliation{Institute for Theoretical Solid State Physics, IFW Dresden, Helmholtzstr.~20, 01069 Dresden,
  Germany}

\author{Inti Sodemann}
\affiliation{Max Planck Institute for the Physics of Complex Systems, N\"{o}thnitzerstr.~38, 01187 Dresden, Germany}

\author{Jeroen van den Brink}
\affiliation{Institute for Theoretical Solid State Physics, IFW Dresden, Helmholtzstr.~20, 01069 Dresden,
  Germany}
\affiliation{Department of Physics, Technical University Dresden, Helmholtzstr.~10, 01062 Dresden, Germany}

\begin{abstract}
Transitions between topologically distinct electronic states have been predicted in different classes of materials and observed in some. A major goal is the identification of measurable properties that directly expose the topological nature of such transitions. Here we focus on the giant-Rashba material bismuth tellurium iodine (BiTeI) which exhibits a pressure-driven phase transition between topological and trivial insulators in three-dimensions. We demonstrate that this transition, which proceeds through an intermediate Weyl semi-metallic state, is accompanied by a giant enhancement of the Berry curvature dipole which can be probed in transport and optoelectronic experiments. From first-principles calculations, we show that the Berrry-dipole --a vector along the polar axis of this material-- has opposite orientations in the trivial and topological insulating phases and peaks at the insulator-to-Weyl critical points, at which the nonlinear Hall conductivity can increase by over two orders of magnitude.
\end{abstract}
\date\today

\maketitle 

A material that is on the verge of a transition between two phases is prone to strongly enhanced responses to small external perturbations.
Such divergent susceptibilities are well-known for critical systems with thermal phase transitions in which the development of long-range order is characterized by a local order parameter.
However, for systems that have a {\it topological} electronic phase transition, in which local order parameter physics is absent, the identification and characterization of measurable signatures of the topological change close to criticality remains largely an open problem.
In this context, we investigate pressure-induced topological electronic phase transitions in the giant Rashba semiconductor BiTeI \cite{ishizaka2011giant}.
This allows us to identify a direct experimental signature of the topological transition: a very strongly enhanced nonlinear Hall conductivity, which is related to a large increase of the Berry curvature dipole moment of the electronic bands close to criticality.

BiTeI is a trivial insulator with a strong Rashba-type spin-orbit coupling and has been predicted to become a strong topological insulator at moderate pressures \cite{Bahramy2012}. These insulating phases are separated by an intermediate Weyl phase \cite{PhysRevB.90.155316, 1367-2630-18-11-113003}.
The ambient pressure crystalline structure has been demonstrated to be stable up to about 9$\,$GPa, which is well above the theoretically expected pressure for the topological phase transition at about 3$\,$GPa \cite{ADMA:ADMA201605965}. Although there are arguments based on optical experiments for the existence of the topological phase transition~\cite{PhysRevLett.111.155701}, the optical properties display a rather smooth evolution up to 9$\,$GPa~\cite{PhysRevLett.112.047402}.
On the other hand, transport experiments present a broad minimum of resistivity at the pressures near where the topological phase transition is expected \cite{ADMA:ADMA201605965,vangennep2017pressure}. Therefore, to our knowledge, there is no experimental evidence of signatures directly associated with the topological nature of the phase transition, such as the appearance of metallic surface states.

A crucial property of BiTeI is the absence of inversion symmetry, which endows the Bloch states with a nonzero Berry curvature. This opens the possibility to analyze the transitions between the topologically distinct electronic phases by means of response functions that are particularly sensitive to the geometry of the Bloch states \cite{sodemann2015quantum,moore2010confinement,deyo2009semiclassical,low2015topological,de2017quantized,PhysRevX.7.041042,luu2018measurement,gavensky2016photo}.
Recent experiments \cite{ma2018observation,kang2018observation} have reported the observation of the time reversal invariant nonlinear Hall effect \cite{sodemann2015quantum} in two-dimensional transition metal dichalcogenides \cite{xu2018electrically,PhysRevB.98.121109,2053-1583-5-4-044001,shi2018berry}. 
Here we demonstrate that such nonlinear Hall effect can also be a useful probe of a topological phase transition and establish BiTeI as a promising platform to experimentally observe this effect in three-dimensions.

\begin{figure*}[t]\center
\includegraphics[width=7.3cm,angle=0,keepaspectratio=true]{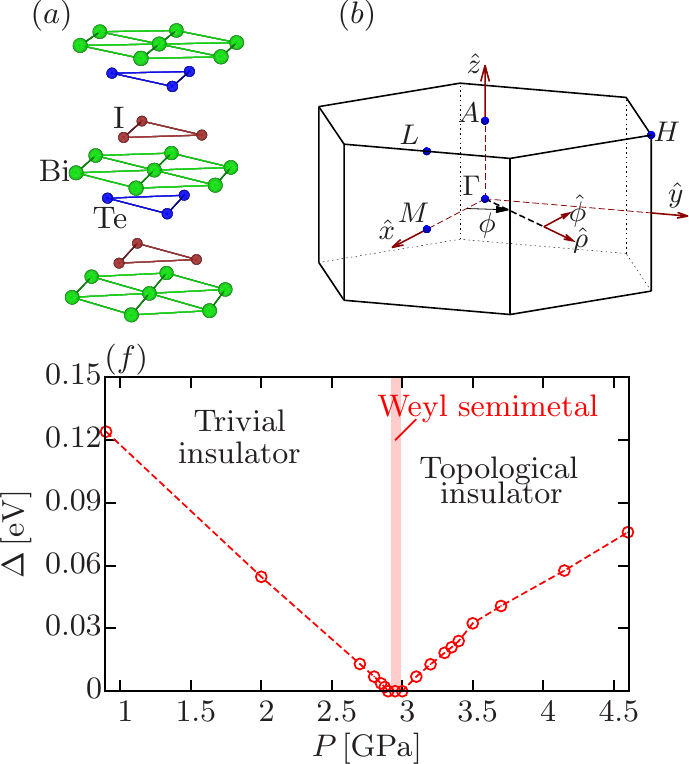}
\includegraphics[width=8.62cm,angle=0,keepaspectratio=true]{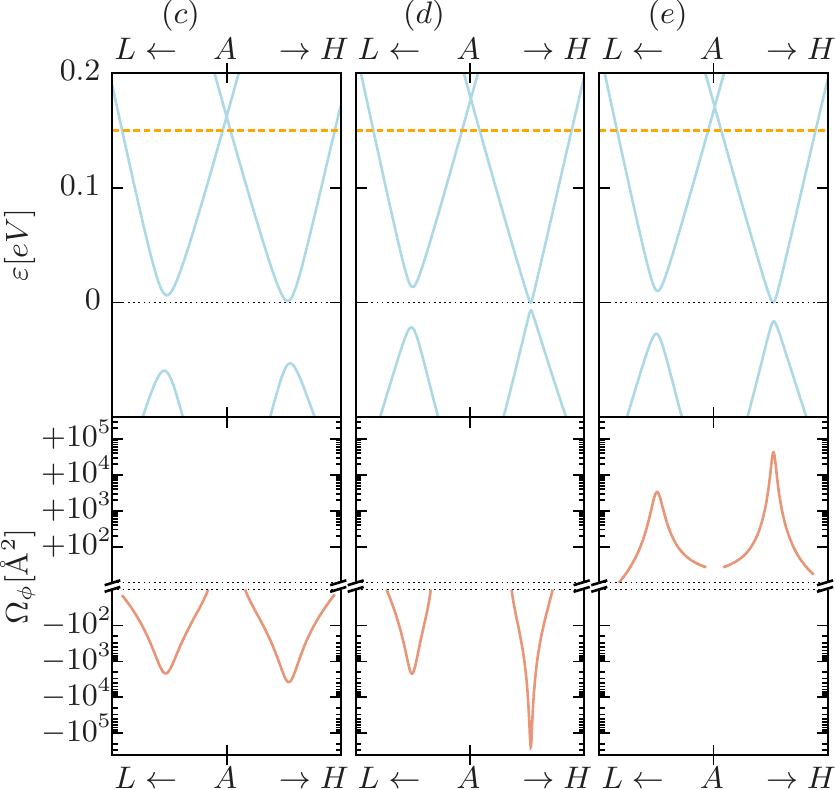}
\caption{$(a)$ Crystal structure of BiTeI. $(b)$ Brillouin zone.
$(c)$, $(d)$ and $(e)$: Band structure (top) and azimuthal component of the total Berry curvature $\Omega_\phi$ (bottom) for pressures $P=2.0$, $2.8$ and $3.2\,$GPa, respectively.
The Berry curvature is that of the states in the conduction band, considering the chemical potential as indicated by the dotted orange line.
$(f)$ Energy gap between valence and conduction states as a function of pressure.
} 
\label{gap}
\end{figure*}

The associated Hall current that is \textit{nonlinear} in the electric field originates from the anomalous velocity caused by the Berry curvature of the electronic bands.
Specifically, in the presence of an electric field $E_c = $Re$\{\mathcal{E}_c$e$^{i\omega t}\}$, the second order response current reads 
$j_a = $Re$\{j_a^{(0)}+j_a^{(2\omega)}$e$^{2i\omega t} \}$,
where $j_a^{(0)} = \chi_{abc} \mathcal{E}_b\mathcal{E}^*_c$ and $j_a^{(2\omega)} = \chi_{abc} \mathcal{E}_b\mathcal{E}_c$.
The nonlinear response function $\chi_{abc}$ effectively measures a \textit{first-order moment} of the Berry curvature over the occupied states, the Berry dipole:

\begin{equation}
\chi_{abc} = -\epsilon_{adc} \frac{\mathrm{e}^3\tau}{2(1+\mathrm{i} \omega\tau)} D_{bd},
\end{equation}
\begin{equation}
D_{bd} = \int \frac{d^3\mathbf{k}}{(2\pi)^3} \sum_n \Omega_{n,d}(\mathbf{k})\Big(-\frac{\partial f_0}{\partial E}\Big)_{E=E_{n\mathbf{k}}}  \frac{\partial E_{n\mathbf{k}}}{\partial k_b}.
\label{dipole}
\end{equation}
Here, $E_\kn$ is the energy dispersion of the $n$-th band, $f_0$ the equilibrium Fermi distribution, $\tau$ the relaxation time and $\Omega_{n,d}$ is the $d$ component of the Berry curvature, $\mathbf{\Omega}_n(\mathbf{k})=\mathbf{\nabla}_\mathbf{k}\times\mathbf{A}_n(\mathbf{k})$, where $\mathbf{A}_n(\mathbf{k})=-\mathrm{i}\langle u_\kn | \partial_\mathbf{k}  u_\kn \rangle$.
In this work, we study the nonlinear Hall conductivity of BiTeI as a function of hydrostatic pressure by computing the Berry dipole $D$ from first-principles.
In addition, we present an analytical description of our mains results based on a low-energy model and finally, we analyze transport and optoelectronic experiments that are expected to exhibit fingerprints of the Berry dipole.

\textit{Topological phase transitions --}
BiTeI is a layered polar compound with space group $P3m1$ (No. 156). Bi, Te and I layers, each having $C_{3v}$ symmetry, are stacked along the $c$ axis breaking inversion symmetry, see Fig. \ref{gap}$(a)$.
To study the evolution under hydrostatic pressure ($P$), we use the experimental values of the lattice parameters $a$ and $c$ reported in Ref. \cite{PhysRevLett.111.155701}.
We also consider additional pressures in between the ones for which experimental data is available, interpolating $a$ and $c$ linearly from the experimental data.
For fixed values of $a$ and $c$, we determine the vertical position of the Te and I atoms by minimization of the total energy.

For each pressure, we perform fully relativistic Density Functional Theory calculations  
\footnote{
We used \textsc{wien2k} \cite{blaha2001wien2k} with the GGA functional, a $24^3$ $k$--mesh and setting $R_{MT}=2.5$ and $K_{max}=8.5R^{-1}_{MT}$.
}
 and compute the Berry curvature of the Bloch states by Wannier interpolation 
\footnote{
	For each pressure we built maximally localized Wannier functions for $p$ orbitals of Bi, Te and I \cite{KUNES20101888,MOSTOFI20142309} and computed $\Omega_{n,d}$ by Wannier interpolation \cite{PhysRevB.74.195118} and $D_{bd}$ as $\int \frac{d^3k}{(2\pi)^3} \sum_n \frac{\partial \Omega_{n,d}}{\partial k_n} f_0(E_\kn)$ using a $1000\times1000\times300$ mesh in the Brillouin-zone region that contains the conducting states in the range of doping analyzed. 
We also performed calculations with \textsc{fplo} \cite{koepernik1999full} based on which we found the same evolution of $\Omega_{n,d}$ with $P$ \cite{Note3}.
} (see Supplementary Material
\footnote{
See Supplementary Material, which includes Refs. \cite{PhysRevB.64.153102,PhysRevB.84.041202,PhysRevB.86.085204, PhysRevB.87.205103,PhysRevB.94.165203}, for: $(i)$ Berry dipole dependence on $\mu$ $(ii)$ discussion of DFT calculations and $(iii)$ formulae associated with the model.
}
).
In three dimensions, the Berry curvature is a vector field defined in the Brillouin zone. We find convenient to measure its components in cylindrical coordinates with the z-axis along the polar axis of BiTeI, as shown in Fig. \ref{gap}$(b)$.
Fig. \ref{gap}$(c-e)$ show for different pressures the band structure and the azimuthal component of the Berry curvature of the conducting bands ($\Omega_\phi$). As we will see later, this is the component that contributes to the Berry dipole in BiTeI.
The energy exhibits the Rashba-like dispersion while $\Omega_\phi$ has extremes at the bottom of the conduction bands.
As $P$ increases, the gap between valence and conduction bands ($\Delta$) is reduced and the extreme of $\Omega_\phi$ along the path $AH$ becomes sharper and achieves larger values.
The gap initially closes at $P_{c1} = 2.9\,$GPa in a point belonging to this path.
Upon further increasing $P$, the Weyl nodes move both in $k_y$ and $k_z$ directions until they annihilate in the mirror plane $AML$ and the gap reopens at $P_{c2} = 3.0\,$GPa. This evolution of the Weyl nodes is in agreement with previous calculations \cite{PhysRevB.90.155316, 1367-2630-18-11-113003}.
For $P>P_{c2}$, increasing the pressure pushes the system deep into the topological insulating phase and the maximum of the Berry curvature is reduced.
Interestingly, due to the band inversion the Berry curvature has opposite sign in the two insulating states.
Fig. \ref{gap}$(f)$ sketches the phase diagram, showing $\Delta$ as a function of $P$.
The pressures in which $\Delta$ vanishes are in excellent agreement with experimental expectations \cite{PhysRevLett.111.155701,ADMA:ADMA201605965}.

\begin{figure}[t]\center
\includegraphics[width=8.4cm,angle=0,keepaspectratio=true]{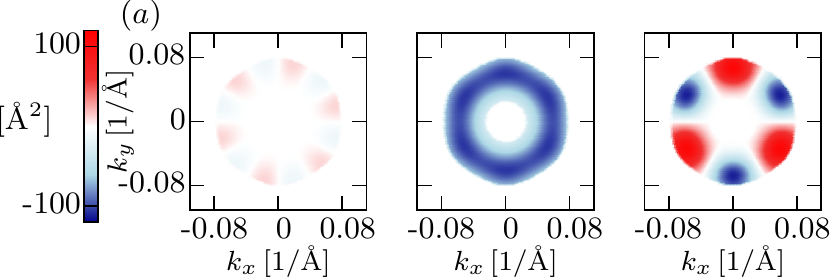}\\
\includegraphics[width=8.4cm,angle=0,keepaspectratio=true]{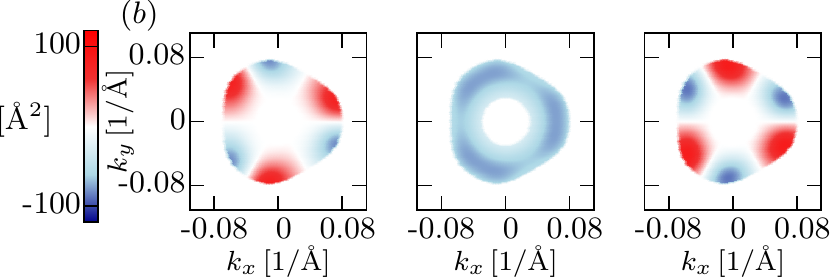}\\
\includegraphics[width=8.2cm,angle=0,keepaspectratio=true]{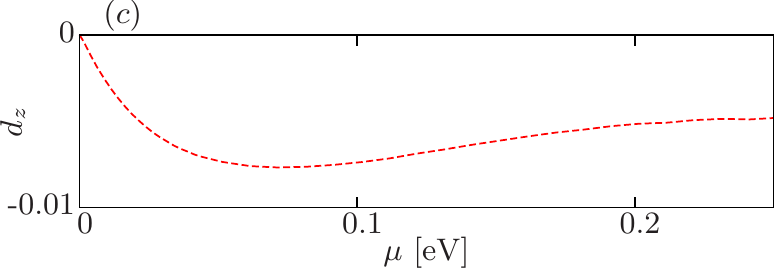}\\
\caption{Berry curvature and Berry dipole at ambient pressure. $(a)$ and $(b)$: Berry curvature of the conducting states of momentum $k_z=\pi/c$ and  $k_z=0.9\pi/c$, respectively. $\Omega_\rho$, $\Omega_\phi$ and $\Omega_z$ are shown from left to right. $\Omega_\rho$ is magnified by a factor of five. $(c)$ Berry dipole as a function of the chemical potential.}
\label{om_cil}
\end{figure}

\textit{Berry curvature dipole --}
We begin by considering how the $C_{3v}$ symmetry constraints the Berry dipole tensor.
Fig. \ref{om_cil}$(a-b)$ show $\bf{\Omega}$ in cylindrical coordinates for particular values of $k_z$.
Since $\bf{\Omega}$ is a pseudo-vector, the components $\Omega_\rho$ and $\Omega_z$ are odd with respect to mirror planes, while $\Omega_\phi$ is even.
As a result, the azimuthal component is seen to swirl around the polar axis in a solenoidal fashion. This circulating pattern is what gives rise to a non-vanishing Berry dipole in BiTeI.
On the other hand, $\Omega_\rho$ and $\Omega_z$ must change sign at the mirrors and their contributions average out to zero.
As explained in Ref. \cite{sodemann2015quantum}, the existence of three mirror planes related by three-fold rotations along the polar axis forces the symmetric part of $D_{bd}$ to vanish.
The remaining antisymmetric part can be described as a vector that lies along the polar axis.
Therefore, we will focus on this vector which can be written in terms of Eq. \ref{dipole} as ${\bf{d}}=d_z \, {\bf \hat{z}}$, where $d_z=(D_{xy}-D_{yx})/2$.

Fig. \ref{om_cil}$(c)$ presents $d_z$ as a function of the chemical potential, $\mu$, at ambient pressure and zero-temperature.
Naturally, without doping ($\mu = 0$), $d_z$ vanishes because of the absence of carriers.
The values obtained at finite doping are moderate as compared to the ones obtained in Ref. \cite{Zhang2018} for different Weyl semimetals, and larger than found in Ref. \cite{PhysRevB.97.035158} for trigonal tellurium.
The maximum in absolute value as a function of $\mu$ 
arises from the competition between the Berry curvature, which is larger at the band bottom, and the growing number of states as the chemical potential increases.
Namely, the maximum corresponds to an electronic density that optimizes the compromise between giving rise to a large Fermi surface and having carriers at the Fermi level with large Berry curvature.
For
a fixed density, hydrostatic pressure can affect this competition by significantly increasing the Berry curvature of the states at the band bottom, as we discuss next.
 
\begin{figure}[t]\center
\includegraphics[width=8.5cm,angle=0,keepaspectratio=true]{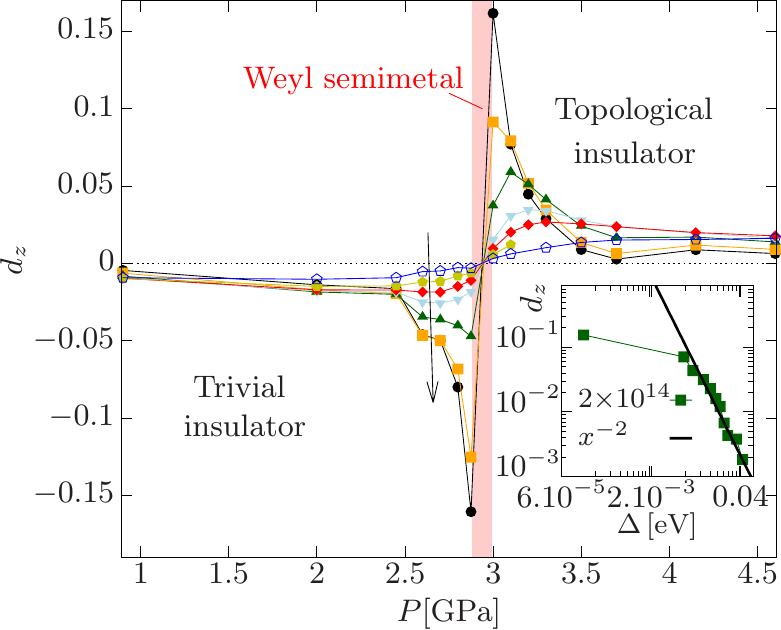}
\caption{Berry dipole as a function of pressure for fixed different values of density. The arrow indicates the direction of decreasing density, the different curves correspond to
$1\times10^{17},\ 5\times10^{16},\, 3\times10^{16},\, 2\times10^{16},\, 1\times10^{16},  3\times10^{15}$ and  $1\times10^{15}\,$cm$^{-3}$.
Inset: Berry dipole as a function of $\Delta$ in the limit of small density in the topological insulating phase.
}
\label{dipvsP}
\end{figure}

Fig. \ref{dipvsP} presents the Berry dipole as a function of pressure for different electronic densities, $n$.
One of the central findings of our work, is that the Berry curvature dipole vector, ${\bf d}=d_z \hat{z}$, reverses its orientation in going from the trivial to the topological insulating phase.
The origin of this reversal can be traced back to the band inversion that causes a change of sign in the Berry curvature (see Fig. \ref{gap}) and, therefore, the measurement of the Berry dipole offers a direct signature of the topological nature of the phase transition.
The evolution with pressure exhibits a noticeable dependence on the amount of doping.
For large $n$, the carriers at the Fermi surface have relatively small Berry curvature and the evolution of $d_z$ is rather smooth.
As the density is reduced, the Fermi surface comes closer to the location of the Weyl nodes in the intervening semimetallic phase, giving rise to a large enhancement of the Berry curvature and its dipole.
In this regime, the Berry dipole not only has a different sign in the two insulating phases but also displays sharp peaks located at the topological phase transitions.
According to our calculations, this dependence on the density is observed up to $n\sim1\times10^{15}\,$cm$^{-3}$.
Reducing further $n$, the decrease in the number of states becomes dominant and, as expected in the limit $\mu \to 0$, $d_z$ decreases its absolute value (see \cite{Note3}).

\textit{Model --} We now introduce a low-energy model to analyze the behaviour of the Berry dipole across the topological phase transitions. 
We first focus on the enhancement obtained when the Weyl phase is approached from the topological insulator side. 
We consider a two-band model for the pair creation or annihilation of Weyl nodes near one of the mirror invariant planes ($ALM$ planes, see Fig. \ref{gap}$(b)$). The valence and conduction bands will touch at some point ${\bf k}_0$ along this plane. The corresponding mirror is the only element of the little group. Choosing $\bf{\hat{x}}$ along the $AL$ line, $\bf{\hat{y}}$ along $LH$ and $\bf{\hat{z}}$ along $\Gamma A$, the mirror acts as:

\begin{equation}
M:k_y\to -k_y, k_{x,z}\to k_{x,z}.
\end{equation}
A simple ``k dot p'' model capturing the symmetries and the pair creation of Weyl nodes is:
\begin{equation}
H=\frac{ \left(k_y^2+\lambda \right)}{2 m} \sigma _y+k_x v_x \sigma _x+k_z v_z \sigma _z + k_x u_x \sigma_0.
\label{eq:H}
\end{equation}
Here $\sigma_{x,y,z}$ are Pauli matrices and $\sigma_0$ is the identity.
The finite tilt $u_x$ term does not change the Berry curvature but affects the Fermi surface shape and is crucial to get a finite Berry dipole.
For $\lambda>0$, this model describes an insulator of gap $\Delta=\frac{\lambda}{|m|}$.
The critical point between this insulator and a Weyl semimetal is at $\lambda=0$. In BiTeI there are six pairs of Weyl points related by discrete $C_{3v}$ operations and time reversal symmetry \cite{PhysRevB.90.155316}. This model captures only one such pair but notice that all the pairs contribute additively to the total Berry dipole \footnote{Weyl points pairs related by $C_3$ naturally contribute equally to $d_z$. 
Those related by time reversal also contribute equally because $\bf{\Omega}$ is odd under time reversal but its curl is even.}.

Assuming we have a finite carrier density $n$ in the conduction band, and far away from the critical point such that $\Delta$ is the largest energy scale and the tilt is small $u_x\ll v_x$, the dipole resulting from the long wave-length Hamiltonian Eq.~\ref{eq:H} is
\begin{equation}
d_z\approx -\frac{n v_z u_x}{m v_x  \Delta^2}.
\label{analitic}
\end{equation}
Thus, as the transition is approached from the topological insulator side, the Berry dipole is strongly enhanced, with an asymptotic behavior as $\sim 1/\Delta^2$ for large $\Delta$.
The inset in Fig. \ref{dipvsP} shows a direct comparison of this analytical result with the full band structure calculations in the low density regime.
 For sufficiently large $\Delta$, the numerical data is indeed consistent with the inverse quadratic behavior, which saturates when $\Delta$ is reduced to values of the order of $\mu$.

The Weyl phase occurs for $\lambda<0$.
The distance between the Weyl nodes is $2 k_0$, with $k_0=\sqrt{|\lambda|}$. 
When the transition is approached from the Weyl phase, the dipole to leading order in $u_x$ and $k_0$, and for $k_0 \to k_F$, is
\begin{equation}
d_z\approx -\frac{ n u_x v_x v_z}{2 k_0 \mu^2 v_y}\left(1-\frac{v_y^2}{v_x^2}\right), 
\label{eq:dip_wsm}
\end{equation}
where $v_y= k_0/m$ is the Fermi velocity along $\bf{\hat{y}}$ near the Weyl point and $k_F$ the Fermi momentum.
We thus see that the dipole is enhanced as $\sim 1/k_0$ as the Weyl points approach each other  
\footnote{Further details of the Weyl semimetal materials will appear in a forthcoming publication}.

Notice that $d_z$ is always finite provided that there is a finite Fermi surface. 
In particular, in the Weyl phase there is no divergence even when the chemical potential is at the nodes 
\footnote{This follows from  Eq. \ref{eq:dip_wsm} since within the Weyl semimetal phase $n \propto \mu^3$. In terms of $\mu$, the Berry dipole in this regime reads $d_z\approx -\frac{ 4\pi \mu u_x}{3 k_0 v_y^2} \left(1-\frac{v_y^2}{v_x^2}\right)$ and vanishes when $\mu \to 0$.}.

A similar analysis can be performed when the transition is approached from the trivial insulator side. The detailed argumentation differs slightly due to a different little group at the point where the conduction and valence bands touch \cite{PhysRevB.90.155316,1367-2630-18-11-113003}, but an analogous model can be developed. The opposite Berry dipole sign in this phase can be understood to arise from the different sign of the effective parameter $m$ which controls the relative ordering in momentum space of the Weyl nodes with opposite chirality. 
This suggests that in other polar materials, like those studied in Ref. \cite{PhysRevB.90.155316}, in which the topological phase transitions occur through a similar Weyl nodes dynamics, or in Ref. \cite{PhysRevB.94.161116}, where the Weyl nodes ordering is reversed with an electric field, 
a change of sign in the Berry dipole can be expected as well.

\textit{Experimental signatures --}
In the following we describe different types of optoelectronic and transport measurements that can be used to probe the Berry dipole in BiTeI.
Throughout this discussion we consider an electric field of frequency smaller than the inter-band optical threshold $\omega \ll \Delta$.
Following Ref. \cite{sodemann2015quantum}, we expect a rectified current at zero frequency and a second harmonic current at $2 \omega$ given by, respectively:
\begin{eqnarray}
{\bf j_0}=\frac{\mathrm{e}^3 \tau}{2(1+ \mathrm{i}   \omega \tau)} {\bf E}^{*} \times (\bf{d}\times {\bf E}).
\end{eqnarray}
\begin{eqnarray}
{\bf j}_{2 \omega}=\frac{\mathrm{e}^3 \tau}{2(1+ \mathrm{i}   \omega \tau)} {\bf E} \times (\bf{d}\times {\bf E}).
\end{eqnarray}
Here the fields ${\bf E}$ are complex vectors to account for nontrivial light polarization.
The formula can also be used to obtain currents in the DC limit by simply taking ${\bf E}$ to be a real vector and multiplying, e.g., the result for ${\bf j_0}$ by a factor of 2 \cite{sodemann2015quantum}.

The electric field polarization can be used to control the direction of ${\bf j_0}$ and ${\bf j_{2\omega}}$ and their relative amplitude.
There are a few special cases of particular interest, the first of which is circularly polarized light.
If the polarization plane is orthogonal to the polar axis, ${\bf E} = E ({\bf \hat{x}} + \mathrm{i}{\bf \hat{y}})$, 
${\bf j}_{2 \omega}$ vanishes and the current reduces to:
\begin{eqnarray}
{\bf j_0}=\frac{\mathrm{e}^3 \tau d_z E^2 \bf{\hat{z}}}{2(1+ \mathrm{i}   \omega \tau)}.
\end{eqnarray}
In contrast, if the polarization plane contains the polar axis, ${\bf E} = E ({\bf \hat{x}} + \mathrm{i}{\bf \hat{z}})$, ${\bf j_0}$ and ${\bf j_{2\omega}}$ have equal amplitude but flow along different directions:
\begin{eqnarray}
{\bf j_0}=\frac{\mathrm{e}^3 \tau d_z E^2 (\bf{\hat{z}}+\bf{\hat{x}})}{2(1+ \mathrm{i}   \omega \tau)}, \, {\bf j}_{2 \omega}=\frac{\mathrm{e}^3 \tau d_z E^2 (\bf{\hat{z}}-\bf{\hat{x}})}{2(1+ \mathrm{i}   \omega \tau)}.
\end{eqnarray}
It is remarkable that the currents flow along orthogonal directions that are $45$ degrees away from the polar axis.
Finally, for linearly polarized light with the electric field at an angle $\theta$ from the polar axis, ${\bf E} = E(\sin\theta {\bf \hat{x}} + \cos\theta {\bf \hat{z}}$), the currents read:
\begin{eqnarray}
{\bf j_0}={\bf j}_{2 \omega}=\frac{\mathrm{e}^3 \tau d_z E^2 \sin\theta}{2(1+ \mathrm{i}   \omega \tau)}(\sin\theta\bf{\hat{z}}-\cos\theta\bf{\hat{x}})
\end{eqnarray}
Notice that the $\bf{\hat{x}}$ and $\bf{\hat{z}}$ current components have different characteristic dependences on $\theta$.

In summary, we have demonstrated that the Berry curvature dipole in BiTeI conveys key information about the topological state of the system.
The Berry dipole vector presents opposite orientations in the topologically distinct insulating phases.
In addition, its magnitude sharply peaks at the phase boundaries between the insulating phases and the intervening Weyl semi-metallic phase.
Optoelectronics and transport measurements, such as the nonlinear Hall effect, are predicted to offer direct experimental evidence of the topological phase transitions in this material under pressure.

We thank Ulrike Nitzsche for technical assistance.  J.v.d.B. acknowledges support from the German Research Foundation (DFG) via SFB 1143.
We are thankful to Snehasish Nandy for discussions and to Liang Fu for suggesting BiTeI as a promising platform to study the Berry curvature dipole. 

\onecolumngrid

\section{Supplemental Material}

\section{Berry curvature as a function of chemical potential for different pressures}

\begin{figure}[h]\center
\includegraphics[width=10cm,angle=0,keepaspectratio=true]{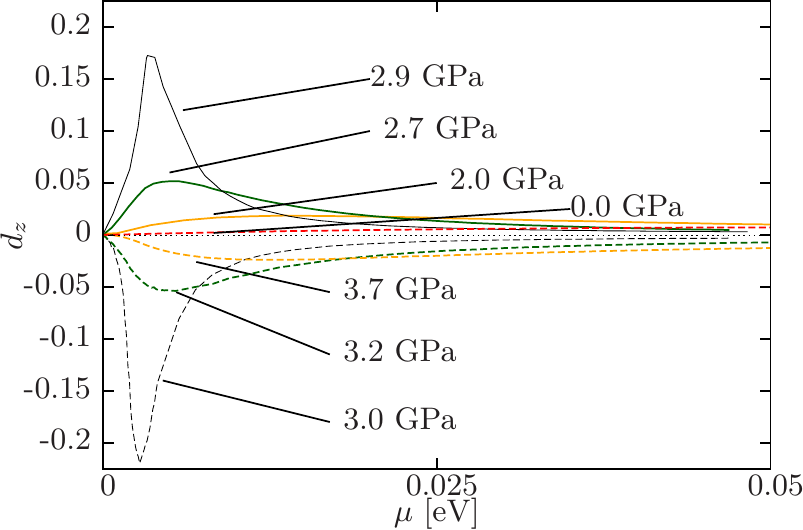}
\caption{Berry dipole as a function of the chemical potential for different values of hydrostatic pressure.}
\label{dip_mu}
\end{figure}

Figure \ref{dip_mu} presents $d_z$ as a function of the chemical potential, $\mu$, at zero-temperature and for different values of pressure, $P$. To understand the existence of a maximum, it is useful to analyze Eq. 2 of the main text. A finite Berry dipole component $D_{yx}$ requires, in addition to sufficiently low symmetry (such that the integral does not vanish), states close to the Fermi level that having a finite band velocity along the $y$ direction have a finite Berry curvature along $x$.  The band velocity also affects implicitly the integral through the density of states at a given energy, so it is convenient to recast this equation at zero temperature as:
\begin{equation}
D_{bd} = \frac{1}{(2\pi)^3}\sum_n\int\limits_{E_{nk}=0} d^2 S v_{n,b}(\mathbf{k})  \Omega_{n,d}(\mathbf{k}),
\label{dipole_fs}
\end{equation}
where $d^2\vec{S}$ is an element of the Fermi surface and $v_{n,b}={\mathbf{\nabla_k}E_{n\mathbf{k}}\cdot \hat{b}}/{|\mathbf{\nabla_k}E_{n\mathbf{k}}|}$. Here it becomes explicit that $D_{yx}$ will be larger as larger portions of the Fermi surface are perpendicular to $y$ while having a large $\Omega_{n,x}$. Since in BiTeI the amplitude of the Berry curvature is maximum for states at the bottom of the conduction band, as $\mu$ is increased the Fermi surface becomes larger but the Berry curvature of the carriers on it diminishes. Therefore, the maximum in the Berry dipole corresponds to an electronic density that optimizes these competing tendencies.

This analysis also anticipates that an interesting behavior is to be expected for BiTeI as a function of hydrostatic pressure, which acts to increase the Berry curvature of the states at the bottom of the conduction band. The pressure has the effect of shifting the maximum of the Berry dipole to lower densities and to significantly increase its value. As the pressure approaches $P_{c1}$, $d_z$ can be increased by over two orders of magnitude as compared to the ambient pressure result. For $P > P_{c2}$, the Berry curvature has a different sign and accordingly does $d_z$. This indicates that $d_z$ goes through a very dramatic change in the narrow range of pressures that correspond to the Weyl semimetal phase. For $P > P_{c2}$, as $P$ is further increased, the extreme of $d_z$ moves back to larger densities and lower values.
The comparison with the analytical result is performed in the limit of small densities (roughly, in Fig. \ref{dip_mu}, this corresponds to values of chemical potential well to the left of the maximum, where $d_z$ becomes larger as $\mu$ increases). In this limit, the Fermi surface is formed by six separated pockets, one associated with each of the parabolic bands near where the Weyl nodes are created at the topological transition. These six pairs of Weyl nodes make the same contribution to the Berry dipole and the model presented in the main text describes one of such pairs.

\section{Methodological aspects of the Density Functional Theory calculations}

\begin{figure}[h]\center
\includegraphics[width=15cm,angle=0,keepaspectratio=true]{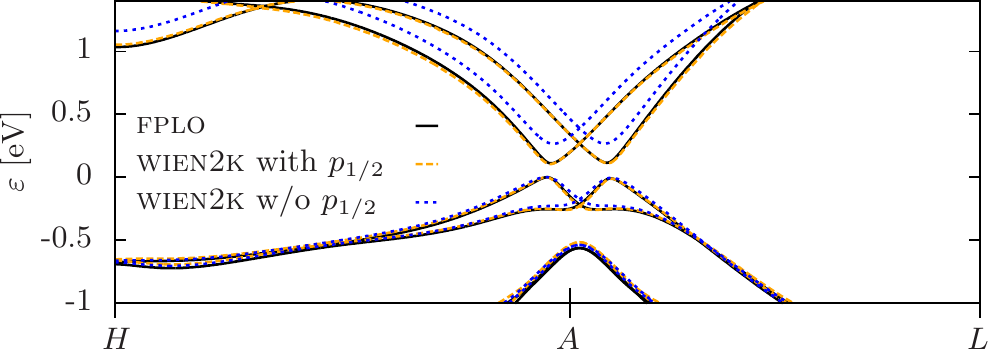}
\caption{Band structure computed with different schemes using the GGA approximation for the exchange and correlation functional.}
\label{bands}
\end{figure}

We performed Density Functional Theory (DFT) calculations of BiTeI using the codes WIEN2K and FPLO. We used the GGA approximation for the exchange and correlation functional. Within WIEN2K, the spin-orbit coupling (SOC) is treated with a second-variational procedure in which the full Hamiltonian including the SOC is diagonalized in the space spanned by scalar relativistic eigenstates.
A well-known aspect of this approach is that the solution may be improved extending this basis with relativistic $p_{1/2}$ local orbitals \cite{PhysRevB.64.153102}.
Our results reproduce those of Ref. \cite{Bahramy2012,PhysRevB.84.041202}, which are based on WIEN2K and show a gap $\Delta\sim0.28\,$eV, only if the $p_{1/2}$ local orbitals are not included in the basis set. When they are included, we find that the gap is reduced to $\Delta\sim0.11\,$eV.
We confirmed this point doing fully relativistic calculations with the FPLO code, in which $p_{1/2}$-like states are included in the basis set by default. Figure \ref{bands} compares the band structure obtained with the different computation schemes used.

\begin{figure}[h]\center
\includegraphics[width=15cm,angle=0,keepaspectratio=true]{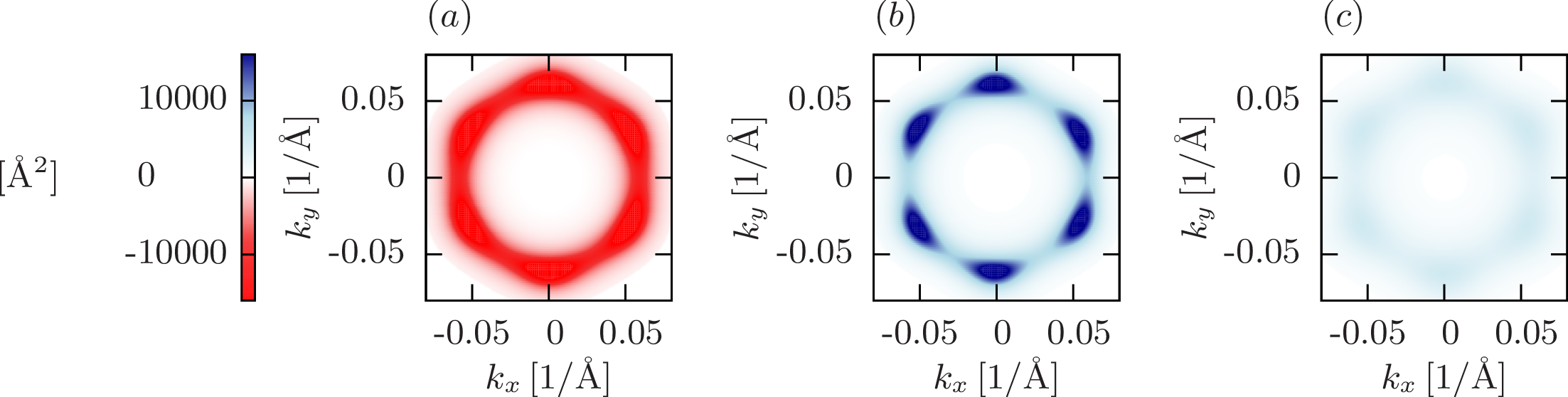}\\
\caption{Azimuthal component of the Berry curvature for different values of volume compression as obtained with FPLO. $(a)$ Trivial insulating phase with $\Delta=0.035\,$eV.
$(b)$: Topological insulating phase with $\Delta=0.05\,$eV. $(c)$: Topological insulating phase $\Delta=0.15\,$eV.}
\label{berrycurvature}
\end{figure}

The experimental value of the gap has been reported to be in the range $[0.26,0.38]\,$eV in Ref. \cite{ishizaka2011giant,PhysRevB.86.085204}.
Detailed \textit{ab initio} studies of $\Delta$ can be found in Ref. \cite{PhysRevB.87.205103,PhysRevB.94.165203}.
While the underestimation of the gap in semiconductors is a well-known feature of semilocal functionals, the fact that the description of the gap is better when the $p_{1/2}$ states are not included may be associated with a fortuitous cancelation of errors. Since this article focuses on the Berry curvature of the conducting states, which is sensitive to the value of $\Delta$, we chose to present in the main text calculations where the $p_{1/2}$ states are not included in order to have a better description of $\Delta$, similar to that reported in Ref. \cite{Bahramy2012,PhysRevB.84.041202}.
This choice only affects the value of pressure at which the transition to the topological nontrivial phases takes place.
Indeed, we have checked that the behavior of the Berry curvature described in the main text is robust to the different computation schemes. This is shown in Fig. \ref{berrycurvature}, which presents the azimuthal component of the Berry curvature for different pressures as obtained with FPLO. The behavior of the Berry curvature across the transition, characterized by an opposite sign in the topologically distinct insulating phases and by an increase in the magnitude when approaching the Weyl semimetal phase, do not depend on the methodology used.  All the calculations indicate that a large increment in the Berry dipole of BiTeI is expected under hydrostatic pressure when $\Delta$ is reduced from its value at ambient pressure, $[0.26,0.38]\,$eV, to 0.

\section{Berry curvature and Berry curvature dipole for the kp model}

The ``k dot p'' model of interest reads:
\begin{equation}
H=\frac{ \left(k_y^2+\lambda \right)}{2 m} \sigma _y+k_x v_x \sigma _x+k_z v_z \sigma_z+ (k_x u_x +k_z u_z) \sigma_0.
\label{eq:H}
\end{equation}
The terms proportional to $\sigma_0$ affect the Fermi surface but do not change the local Berry curvature.
For computing each of the Berry curvature components the problem can be seen as effectively 2D.
For instance, let us first focus on $\Omega_z=\partial_{kx}A_y-\partial_{ky}A_x$.
For its calculation we can consider the Hamiltonian:
\begin{equation}
H=v_x ( k_x \sigma_x+k_y'\sigma_y) + m_z \sigma _z + C ,
\label{eq:H}
\end{equation}
where $k_y'=\frac{ \left(k_y^2+\lambda \right)}{2 m v_x}$, $m_z = k_z v_z$ and $C$ is a constant.
Then, noticing that
\begin{eqnarray}
A_y &=& -i \bra{u} \partial_{ky} \ket{u} = \frac{k_y}{m v_x}(\partial_{kx}A_y'-\partial_{ky'}A_x) = \frac{k_y}{m v_x} A_y',\\
\Omega_z &=& \partial_{kx}A_y-\partial_{ky}A_x = \frac{k_y}{m v_x}(\partial_{kx}A_y'-\partial_{ky'}A_x) = \frac{k_y}{m v_x} \Omega_z',
\end{eqnarray}
we can use the classic result of the Berry curvature for a 2D a massive Dirac fermion (see e.g. Eq.(17) from Ref. \cite{sodemann2015quantum}) and obtain:
\begin{equation}
\Omega_z = \text{sign}(\mu)\frac{v_z v_x k_z k_y}{2mr^{3/2}}
\label{omz}
\end{equation}
where $r=r(k_x,k_y,k_z)=(v_x k_x)^2 + (v_z k_z)^2 + (\frac{1}{2m}(k_y^2+\lambda))^2$.
Next, for computing $\Omega_x$ we can consider:
\begin{eqnarray}
	\Omega_x &=& -(\partial_{kz}A_y-\partial_{ky}A_z), \\
	H &=& m_x \sigma_x + v_z (k_z \sigma_z + k_y' \sigma_y) + C,
\end{eqnarray}
where $k_y'=\frac{ \left(k_y^2+\lambda \right)}{2 m v_z}$ and $m_x = k_x v_x$.
Thus, we see that the calculation of $-\Omega_x$ is equivalent to the one above after exchanging $x$ and $z$ labels
and performing a unitary transformation that sends $\sigma_y \to \sigma_y$, $\sigma_x \to -\sigma_z$ and $\sigma_z$ to $\sigma_x$, which would give an extra global minus sign in the Berry curvature formula.
We obtain
\begin{equation}
\Omega_x = \text{sign}(\mu) \frac{v_z v_x k_x k_y}{2mr^{3/2}}.
\end{equation}
For $\Omega_y$, we can again follow the same kind of rationale and we obtain:
\begin{equation}
\Omega_y = \text{sign}(\mu) \frac{v_z v_x (k_y^2+\lambda)}{4mr^{3/2}}.
\end{equation}
Now we can compute the antisymmetric dipole along the $z$ axis, which is controlled by the integral of the curl
of the Berry curvature:
\begin{equation}
d_z(k_x,k_y,kz) = \partial_{kx}\Omega_y-\partial_{ky}\Omega_x = \frac{\text{sign}(\mu)k_x v_x v_z }{2m r^{3/2}} \Big[\frac{3(k_y^2+\lambda)}{2r}\Big(\frac{k_y^2}{m^2}-v_x^2\Big) -1\Big].
\label{dk}
\end{equation}
As we can see the Berry dipole is even with respect to $k_y$ and $k_z$ but odd with $k_x$.
Thus the tilt controlled by $u_x$ is crucial to get a finite value upon integration.
The formulas for the Berry dipole presented in the main text for different limits are obtained after integration of Eq. \ref{dk} for all states up to the Fermi energy.
\twocolumngrid

\bibliography{ref}

\begin{thebibliography}{41}%
\makeatletter
\providecommand \@ifxundefined [1]{%
 \@ifx{#1\undefined}
}%
\providecommand \@ifnum [1]{%
 \ifnum #1\expandafter \@firstoftwo
 \else \expandafter \@secondoftwo
 \fi
}%
\providecommand \@ifx [1]{%
 \ifx #1\expandafter \@firstoftwo
 \else \expandafter \@secondoftwo
 \fi
}%
\providecommand \natexlab [1]{#1}%
\providecommand \enquote  [1]{``#1''}%
\providecommand \bibnamefont  [1]{#1}%
\providecommand \bibfnamefont [1]{#1}%
\providecommand \citenamefont [1]{#1}%
\providecommand \href@noop [0]{\@secondoftwo}%
\providecommand \href [0]{\begingroup \@sanitize@url \@href}%
\providecommand \@href[1]{\@@startlink{#1}\@@href}%
\providecommand \@@href[1]{\endgroup#1\@@endlink}%
\providecommand \@sanitize@url [0]{\catcode `\\12\catcode `\$12\catcode
  `\&12\catcode `\#12\catcode `\^12\catcode `\_12\catcode `\%12\relax}%
\providecommand \@@startlink[1]{}%
\providecommand \@@endlink[0]{}%
\providecommand \url  [0]{\begingroup\@sanitize@url \@url }%
\providecommand \@url [1]{\endgroup\@href {#1}{\urlprefix }}%
\providecommand \urlprefix  [0]{URL }%
\providecommand \Eprint [0]{\href }%
\providecommand \doibase [0]{http://dx.doi.org/}%
\providecommand \selectlanguage [0]{\@gobble}%
\providecommand \bibinfo  [0]{\@secondoftwo}%
\providecommand \bibfield  [0]{\@secondoftwo}%
\providecommand \translation [1]{[#1]}%
\providecommand \BibitemOpen [0]{}%
\providecommand \bibitemStop [0]{}%
\providecommand \bibitemNoStop [0]{.\EOS\space}%
\providecommand \EOS [0]{\spacefactor3000\relax}%
\providecommand \BibitemShut  [1]{\csname bibitem#1\endcsname}%
\let\auto@bib@innerbib\@empty
\bibitem [{\citenamefont {Ishizaka}\ \emph {et~al.}(2011)\citenamefont
  {Ishizaka}, \citenamefont {Bahramy}, \citenamefont {Murakawa}, \citenamefont
  {Sakano}, \citenamefont {Shimojima}, \citenamefont {Sonobe}, \citenamefont
  {Koizumi}, \citenamefont {Shin}, \citenamefont {Miyahara}, \citenamefont
  {Kimura} \emph {et~al.}}]{ishizaka2011giant}%
  \BibitemOpen
  \bibfield  {author} {\bibinfo {author} {\bibfnamefont {K}~\bibnamefont
  {Ishizaka}}, \bibinfo {author} {\bibfnamefont {MS}~\bibnamefont {Bahramy}},
  \bibinfo {author} {\bibfnamefont {H}~\bibnamefont {Murakawa}}, \bibinfo
  {author} {\bibfnamefont {M}~\bibnamefont {Sakano}}, \bibinfo {author}
  {\bibfnamefont {T}~\bibnamefont {Shimojima}}, \bibinfo {author}
  {\bibfnamefont {T}~\bibnamefont {Sonobe}}, \bibinfo {author} {\bibfnamefont
  {K}~\bibnamefont {Koizumi}}, \bibinfo {author} {\bibfnamefont
  {S}~\bibnamefont {Shin}}, \bibinfo {author} {\bibfnamefont {H}~\bibnamefont
  {Miyahara}}, \bibinfo {author} {\bibfnamefont {A}~\bibnamefont {Kimura}},
  \emph {et~al.},\ }\bibfield  {title} {\enquote {\bibinfo {title} {Giant
  {Rashba}-type spin splitting in bulk {BiTeI}},}\ }\href@noop {} {\bibfield
  {journal} {\bibinfo  {journal} {Nature materials}\ }\textbf {\bibinfo
  {volume} {10}},\ \bibinfo {pages} {521} (\bibinfo {year} {2011})}\BibitemShut
  {NoStop}%
\bibitem [{\citenamefont {Bahramy}\ \emph {et~al.}(2012)\citenamefont
  {Bahramy}, \citenamefont {Yang}, \citenamefont {Arita},\ and\ \citenamefont
  {Nagaosa}}]{Bahramy2012}%
  \BibitemOpen
  \bibfield  {author} {\bibinfo {author} {\bibfnamefont {MS}~\bibnamefont
  {Bahramy}}, \bibinfo {author} {\bibfnamefont {B-J}\ \bibnamefont {Yang}},
  \bibinfo {author} {\bibfnamefont {R}~\bibnamefont {Arita}}, \ and\ \bibinfo
  {author} {\bibfnamefont {N}~\bibnamefont {Nagaosa}},\ }\bibfield  {title}
  {\enquote {\bibinfo {title} {Emergence of non-centrosymmetric topological
  insulating phase in {BiTeI} under pressure},}\ }\href@noop {} {\bibfield
  {journal} {\bibinfo  {journal} {Nature Communications}\ }\textbf {\bibinfo
  {volume} {3}},\ \bibinfo {pages} {679} (\bibinfo {year} {2012})}\BibitemShut
  {NoStop}%
\bibitem [{\citenamefont {Liu}\ and\ \citenamefont
  {Vanderbilt}(2014)}]{PhysRevB.90.155316}%
  \BibitemOpen
  \bibfield  {author} {\bibinfo {author} {\bibfnamefont {Jianpeng}\
  \bibnamefont {Liu}}\ and\ \bibinfo {author} {\bibfnamefont {David}\
  \bibnamefont {Vanderbilt}},\ }\bibfield  {title} {\enquote {\bibinfo {title}
  {Weyl semimetals from noncentrosymmetric topological insulators},}\ }\href
  {\doibase 10.1103/PhysRevB.90.155316} {\bibfield  {journal} {\bibinfo
  {journal} {Phys. Rev. B}\ }\textbf {\bibinfo {volume} {90}},\ \bibinfo
  {pages} {155316} (\bibinfo {year} {2014})}\BibitemShut {NoStop}%
\bibitem [{\citenamefont {Rusinov}\ \emph {et~al.}(2016)\citenamefont
  {Rusinov}, \citenamefont {Menshchikova}, \citenamefont {Sklyadneva},
  \citenamefont {Heid}, \citenamefont {Bohnen},\ and\ \citenamefont
  {Chulkov}}]{1367-2630-18-11-113003}%
  \BibitemOpen
  \bibfield  {author} {\bibinfo {author} {\bibfnamefont {I~P}\ \bibnamefont
  {Rusinov}}, \bibinfo {author} {\bibfnamefont {T~V}\ \bibnamefont
  {Menshchikova}}, \bibinfo {author} {\bibfnamefont {I~Yu}\ \bibnamefont
  {Sklyadneva}}, \bibinfo {author} {\bibfnamefont {R}~\bibnamefont {Heid}},
  \bibinfo {author} {\bibfnamefont {K-P}\ \bibnamefont {Bohnen}}, \ and\
  \bibinfo {author} {\bibfnamefont {E~V}\ \bibnamefont {Chulkov}},\ }\bibfield
  {title} {\enquote {\bibinfo {title} {Pressure effects on crystal and
  electronic structure of bismuth tellurohalides},}\ }\href
  {http://stacks.iop.org/1367-2630/18/i=11/a=113003} {\bibfield  {journal}
  {\bibinfo  {journal} {New Journal of Physics}\ }\textbf {\bibinfo {volume}
  {18}},\ \bibinfo {pages} {113003} (\bibinfo {year} {2016})}\BibitemShut
  {NoStop}%
\bibitem [{\citenamefont {Qi}\ \emph {et~al.}(2017)\citenamefont {Qi},
  \citenamefont {Shi}, \citenamefont {Naumov}, \citenamefont {Kumar},
  \citenamefont {Sankar}, \citenamefont {Schnelle}, \citenamefont {Shekhar},
  \citenamefont {Chou}, \citenamefont {Felser}, \citenamefont {Yan},\ and\
  \citenamefont {Medvedev}}]{ADMA:ADMA201605965}%
  \BibitemOpen
  \bibfield  {author} {\bibinfo {author} {\bibfnamefont {Yanpeng}\ \bibnamefont
  {Qi}}, \bibinfo {author} {\bibfnamefont {Wujun}\ \bibnamefont {Shi}},
  \bibinfo {author} {\bibfnamefont {Pavel~G.}\ \bibnamefont {Naumov}}, \bibinfo
  {author} {\bibfnamefont {Nitesh}\ \bibnamefont {Kumar}}, \bibinfo {author}
  {\bibfnamefont {Raman}\ \bibnamefont {Sankar}}, \bibinfo {author}
  {\bibfnamefont {Walter}\ \bibnamefont {Schnelle}}, \bibinfo {author}
  {\bibfnamefont {Chandra}\ \bibnamefont {Shekhar}}, \bibinfo {author}
  {\bibfnamefont {Fang-Cheng}\ \bibnamefont {Chou}}, \bibinfo {author}
  {\bibfnamefont {Claudia}\ \bibnamefont {Felser}}, \bibinfo {author}
  {\bibfnamefont {Binghai}\ \bibnamefont {Yan}}, \ and\ \bibinfo {author}
  {\bibfnamefont {Sergey~A.}\ \bibnamefont {Medvedev}},\ }\bibfield  {title}
  {\enquote {\bibinfo {title} {Topological quantum phase transition and
  superconductivity induced by pressure in the bismuth tellurohalide
  {BiTeI}},}\ }\href {\doibase 10.1002/adma.201605965} {\bibfield  {journal}
  {\bibinfo  {journal} {Advanced Materials}\ }\textbf {\bibinfo {volume}
  {29}},\ \bibinfo {pages} {1605965--n/a} (\bibinfo {year} {2017})},\ \bibinfo
  {note} {1605965}\BibitemShut {NoStop}%
\bibitem [{\citenamefont {Xi}\ \emph {et~al.}(2013)\citenamefont {Xi},
  \citenamefont {Ma}, \citenamefont {Liu}, \citenamefont {Chen}, \citenamefont
  {Ku}, \citenamefont {Berger}, \citenamefont {Martin}, \citenamefont
  {Tanner},\ and\ \citenamefont {Carr}}]{PhysRevLett.111.155701}%
  \BibitemOpen
  \bibfield  {author} {\bibinfo {author} {\bibfnamefont {Xiaoxiang}\
  \bibnamefont {Xi}}, \bibinfo {author} {\bibfnamefont {Chunli}\ \bibnamefont
  {Ma}}, \bibinfo {author} {\bibfnamefont {Zhenxian}\ \bibnamefont {Liu}},
  \bibinfo {author} {\bibfnamefont {Zhiqiang}\ \bibnamefont {Chen}}, \bibinfo
  {author} {\bibfnamefont {Wei}\ \bibnamefont {Ku}}, \bibinfo {author}
  {\bibfnamefont {H.}~\bibnamefont {Berger}}, \bibinfo {author} {\bibfnamefont
  {C.}~\bibnamefont {Martin}}, \bibinfo {author} {\bibfnamefont {D.~B.}\
  \bibnamefont {Tanner}}, \ and\ \bibinfo {author} {\bibfnamefont {G.~L.}\
  \bibnamefont {Carr}},\ }\bibfield  {title} {\enquote {\bibinfo {title}
  {Signatures of a pressure-induced topological quantum phase transition in
  {BiTeI}},}\ }\href {\doibase 10.1103/PhysRevLett.111.155701} {\bibfield
  {journal} {\bibinfo  {journal} {Phys. Rev. Lett.}\ }\textbf {\bibinfo
  {volume} {111}},\ \bibinfo {pages} {155701} (\bibinfo {year}
  {2013})}\BibitemShut {NoStop}%
\bibitem [{\citenamefont {Tran}\ \emph {et~al.}(2014)\citenamefont {Tran},
  \citenamefont {Levallois}, \citenamefont {Lerch}, \citenamefont {Teyssier},
  \citenamefont {Kuzmenko}, \citenamefont {Aut\`es}, \citenamefont {Yazyev},
  \citenamefont {Ubaldini}, \citenamefont {Giannini}, \citenamefont {van~der
  Marel},\ and\ \citenamefont {Akrap}}]{PhysRevLett.112.047402}%
  \BibitemOpen
  \bibfield  {author} {\bibinfo {author} {\bibfnamefont {M.~K.}\ \bibnamefont
  {Tran}}, \bibinfo {author} {\bibfnamefont {J.}~\bibnamefont {Levallois}},
  \bibinfo {author} {\bibfnamefont {P.}~\bibnamefont {Lerch}}, \bibinfo
  {author} {\bibfnamefont {J.}~\bibnamefont {Teyssier}}, \bibinfo {author}
  {\bibfnamefont {A.~B.}\ \bibnamefont {Kuzmenko}}, \bibinfo {author}
  {\bibfnamefont {G.}~\bibnamefont {Aut\`es}}, \bibinfo {author} {\bibfnamefont
  {O.~V.}\ \bibnamefont {Yazyev}}, \bibinfo {author} {\bibfnamefont
  {A.}~\bibnamefont {Ubaldini}}, \bibinfo {author} {\bibfnamefont
  {E.}~\bibnamefont {Giannini}}, \bibinfo {author} {\bibfnamefont
  {D.}~\bibnamefont {van~der Marel}}, \ and\ \bibinfo {author} {\bibfnamefont
  {A.}~\bibnamefont {Akrap}},\ }\bibfield  {title} {\enquote {\bibinfo {title}
  {Infrared- and raman-spectroscopy measurements of a transition in the crystal
  structure and a closing of the energy gap of {BiTeI} under pressure},}\
  }\href {\doibase 10.1103/PhysRevLett.112.047402} {\bibfield  {journal}
  {\bibinfo  {journal} {Phys. Rev. Lett.}\ }\textbf {\bibinfo {volume} {112}},\
  \bibinfo {pages} {047402} (\bibinfo {year} {2014})}\BibitemShut {NoStop}%
\bibitem [{\citenamefont {VanGennep}\ \emph {et~al.}(2017)\citenamefont
  {VanGennep}, \citenamefont {Linscheid}, \citenamefont {Jackson},
  \citenamefont {Weir}, \citenamefont {Vohra}, \citenamefont {Berger},
  \citenamefont {Stewart}, \citenamefont {Hennig}, \citenamefont {Hirschfeld},\
  and\ \citenamefont {Hamlin}}]{vangennep2017pressure}%
  \BibitemOpen
  \bibfield  {author} {\bibinfo {author} {\bibfnamefont {Derrick}\ \bibnamefont
  {VanGennep}}, \bibinfo {author} {\bibfnamefont {Andreas}\ \bibnamefont
  {Linscheid}}, \bibinfo {author} {\bibfnamefont {DE}~\bibnamefont {Jackson}},
  \bibinfo {author} {\bibfnamefont {ST}~\bibnamefont {Weir}}, \bibinfo {author}
  {\bibfnamefont {YK}~\bibnamefont {Vohra}}, \bibinfo {author} {\bibfnamefont
  {Helmuth}\ \bibnamefont {Berger}}, \bibinfo {author} {\bibfnamefont
  {GR}~\bibnamefont {Stewart}}, \bibinfo {author} {\bibfnamefont
  {RG}~\bibnamefont {Hennig}}, \bibinfo {author} {\bibfnamefont
  {PJ}~\bibnamefont {Hirschfeld}}, \ and\ \bibinfo {author} {\bibfnamefont
  {JJ}~\bibnamefont {Hamlin}},\ }\bibfield  {title} {\enquote {\bibinfo {title}
  {Pressure-induced superconductivity in the giant {Rashba system BiTeI}},}\
  }\href@noop {} {\bibfield  {journal} {\bibinfo  {journal} {Journal of
  Physics: Condensed Matter}\ }\textbf {\bibinfo {volume} {29}},\ \bibinfo
  {pages} {09LT02} (\bibinfo {year} {2017})}\BibitemShut {NoStop}%
\bibitem [{\citenamefont {Sodemann}\ and\ \citenamefont
  {Fu}(2015)}]{sodemann2015quantum}%
  \BibitemOpen
  \bibfield  {author} {\bibinfo {author} {\bibfnamefont {Inti}\ \bibnamefont
  {Sodemann}}\ and\ \bibinfo {author} {\bibfnamefont {Liang}\ \bibnamefont
  {Fu}},\ }\bibfield  {title} {\enquote {\bibinfo {title} {Quantum nonlinear
  {Hall} effect induced by {Berry} curvature dipole in time-reversal invariant
  materials},}\ }\href@noop {} {\bibfield  {journal} {\bibinfo  {journal}
  {Physical Review Letters}\ }\textbf {\bibinfo {volume} {115}},\ \bibinfo
  {pages} {216806} (\bibinfo {year} {2015})}\BibitemShut {NoStop}%
\bibitem [{\citenamefont {Moore}\ and\ \citenamefont
  {Orenstein}(2010)}]{moore2010confinement}%
  \BibitemOpen
  \bibfield  {author} {\bibinfo {author} {\bibfnamefont {Joel~E}\ \bibnamefont
  {Moore}}\ and\ \bibinfo {author} {\bibfnamefont {J}~\bibnamefont
  {Orenstein}},\ }\bibfield  {title} {\enquote {\bibinfo {title}
  {Confinement-induced {Berry} phase and helicity-dependent photocurrents},}\
  }\href@noop {} {\bibfield  {journal} {\bibinfo  {journal} {Physical Review
  Letters}\ }\textbf {\bibinfo {volume} {105}},\ \bibinfo {pages} {026805}
  (\bibinfo {year} {2010})}\BibitemShut {NoStop}%
\bibitem [{\citenamefont {Deyo}\ \emph {et~al.}(2009)\citenamefont {Deyo},
  \citenamefont {Golub}, \citenamefont {Ivchenko},\ and\ \citenamefont
  {Spivak}}]{deyo2009semiclassical}%
  \BibitemOpen
  \bibfield  {author} {\bibinfo {author} {\bibfnamefont {E}~\bibnamefont
  {Deyo}}, \bibinfo {author} {\bibfnamefont {LE}~\bibnamefont {Golub}},
  \bibinfo {author} {\bibfnamefont {EL}~\bibnamefont {Ivchenko}}, \ and\
  \bibinfo {author} {\bibfnamefont {B}~\bibnamefont {Spivak}},\ }\bibfield
  {title} {\enquote {\bibinfo {title} {Semiclassical theory of the
  photogalvanic effect in non-centrosymmetric systems},}\ }\href@noop {}
  {\bibfield  {journal} {\bibinfo  {journal} {Preprint at
  https://arxiv.org/abs/0904.1917v1}\ } (\bibinfo {year} {2009})}\BibitemShut
  {NoStop}%
\bibitem [{\citenamefont {Low}\ \emph {et~al.}(2015)\citenamefont {Low},
  \citenamefont {Jiang},\ and\ \citenamefont {Guinea}}]{low2015topological}%
  \BibitemOpen
  \bibfield  {author} {\bibinfo {author} {\bibfnamefont {Tony}\ \bibnamefont
  {Low}}, \bibinfo {author} {\bibfnamefont {Yongjin}\ \bibnamefont {Jiang}}, \
  and\ \bibinfo {author} {\bibfnamefont {Francisco}\ \bibnamefont {Guinea}},\
  }\bibfield  {title} {\enquote {\bibinfo {title} {Topological currents in
  black phosphorus with broken inversion symmetry},}\ }\href@noop {} {\bibfield
   {journal} {\bibinfo  {journal} {Physical Review B}\ }\textbf {\bibinfo
  {volume} {92}},\ \bibinfo {pages} {235447} (\bibinfo {year}
  {2015})}\BibitemShut {NoStop}%
\bibitem [{\citenamefont {de~Juan}\ \emph {et~al.}(2017)\citenamefont
  {de~Juan}, \citenamefont {Grushin}, \citenamefont {Morimoto},\ and\
  \citenamefont {Moore}}]{de2017quantized}%
  \BibitemOpen
  \bibfield  {author} {\bibinfo {author} {\bibfnamefont {Fernando}\
  \bibnamefont {de~Juan}}, \bibinfo {author} {\bibfnamefont {Adolfo~G}\
  \bibnamefont {Grushin}}, \bibinfo {author} {\bibfnamefont {Takahiro}\
  \bibnamefont {Morimoto}}, \ and\ \bibinfo {author} {\bibfnamefont {Joel~E}\
  \bibnamefont {Moore}},\ }\bibfield  {title} {\enquote {\bibinfo {title}
  {Quantized circular photogalvanic effect in {Weyl} semimetals},}\ }\href@noop
  {} {\bibfield  {journal} {\bibinfo  {journal} {Nature Communications}\
  }\textbf {\bibinfo {volume} {8}},\ \bibinfo {pages} {15995} (\bibinfo {year}
  {2017})}\BibitemShut {NoStop}%
\bibitem [{\citenamefont {Banks}\ \emph {et~al.}(2017)\citenamefont {Banks},
  \citenamefont {Wu}, \citenamefont {Valovcin}, \citenamefont {Mack},
  \citenamefont {Gossard}, \citenamefont {Pfeiffer}, \citenamefont {Liu},\ and\
  \citenamefont {Sherwin}}]{PhysRevX.7.041042}%
  \BibitemOpen
  \bibfield  {author} {\bibinfo {author} {\bibfnamefont {Hunter~B.}\
  \bibnamefont {Banks}}, \bibinfo {author} {\bibfnamefont {Qile}\ \bibnamefont
  {Wu}}, \bibinfo {author} {\bibfnamefont {Darren~C.}\ \bibnamefont
  {Valovcin}}, \bibinfo {author} {\bibfnamefont {Shawn}\ \bibnamefont {Mack}},
  \bibinfo {author} {\bibfnamefont {Arthur~C.}\ \bibnamefont {Gossard}},
  \bibinfo {author} {\bibfnamefont {Loren}\ \bibnamefont {Pfeiffer}}, \bibinfo
  {author} {\bibfnamefont {Ren-Bao}\ \bibnamefont {Liu}}, \ and\ \bibinfo
  {author} {\bibfnamefont {Mark~S.}\ \bibnamefont {Sherwin}},\ }\bibfield
  {title} {\enquote {\bibinfo {title} {Dynamical birefringence: Electron-hole
  recollisions as probes of {Berry} curvature},}\ }\href {\doibase
  10.1103/PhysRevX.7.041042} {\bibfield  {journal} {\bibinfo  {journal} {Phys.
  Rev. X}\ }\textbf {\bibinfo {volume} {7}},\ \bibinfo {pages} {041042}
  (\bibinfo {year} {2017})}\BibitemShut {NoStop}%
\bibitem [{\citenamefont {Luu}\ and\ \citenamefont
  {W{\"o}rner}(2018)}]{luu2018measurement}%
  \BibitemOpen
  \bibfield  {author} {\bibinfo {author} {\bibfnamefont {Tran~Trung}\
  \bibnamefont {Luu}}\ and\ \bibinfo {author} {\bibfnamefont {Hans~Jakob}\
  \bibnamefont {W{\"o}rner}},\ }\bibfield  {title} {\enquote {\bibinfo {title}
  {Measurement of the {Berry} curvature of solids using high-harmonic
  spectroscopy},}\ }\href@noop {} {\bibfield  {journal} {\bibinfo  {journal}
  {Nature Communications}\ }\textbf {\bibinfo {volume} {9}},\ \bibinfo {pages}
  {916} (\bibinfo {year} {2018})}\BibitemShut {NoStop}%
\bibitem [{\citenamefont {Gavensky}\ \emph {et~al.}(2016)\citenamefont
  {Gavensky}, \citenamefont {Usaj},\ and\ \citenamefont
  {Balseiro}}]{gavensky2016photo}%
  \BibitemOpen
  \bibfield  {author} {\bibinfo {author} {\bibfnamefont {Lucila~Peralta}\
  \bibnamefont {Gavensky}}, \bibinfo {author} {\bibfnamefont {Gonzalo}\
  \bibnamefont {Usaj}}, \ and\ \bibinfo {author} {\bibfnamefont
  {CA}~\bibnamefont {Balseiro}},\ }\bibfield  {title} {\enquote {\bibinfo
  {title} {Photo-electrons unveil topological transitions in graphene-like
  systems},}\ }\href@noop {} {\bibfield  {journal} {\bibinfo  {journal}
  {Scientific Reports}\ }\textbf {\bibinfo {volume} {6}},\ \bibinfo {pages}
  {36577} (\bibinfo {year} {2016})}\BibitemShut {NoStop}%
\bibitem [{\citenamefont {Ma}\ \emph {et~al.}(2018)\citenamefont {Ma},
  \citenamefont {Xu}, \citenamefont {Shen}, \citenamefont {Macneill},
  \citenamefont {Fatemi}, \citenamefont {Valdivia}, \citenamefont {Wu},
  \citenamefont {Chang}, \citenamefont {Du}, \citenamefont {Hsu} \emph
  {et~al.}}]{ma2018observation}%
  \BibitemOpen
  \bibfield  {author} {\bibinfo {author} {\bibfnamefont {Qiong}\ \bibnamefont
  {Ma}}, \bibinfo {author} {\bibfnamefont {Su-Yang}\ \bibnamefont {Xu}},
  \bibinfo {author} {\bibfnamefont {Huitao}\ \bibnamefont {Shen}}, \bibinfo
  {author} {\bibfnamefont {David}\ \bibnamefont {Macneill}}, \bibinfo {author}
  {\bibfnamefont {Valla}\ \bibnamefont {Fatemi}}, \bibinfo {author}
  {\bibfnamefont {Andres M~Mier}\ \bibnamefont {Valdivia}}, \bibinfo {author}
  {\bibfnamefont {Sanfeng}\ \bibnamefont {Wu}}, \bibinfo {author}
  {\bibfnamefont {Tay-Rong}\ \bibnamefont {Chang}}, \bibinfo {author}
  {\bibfnamefont {Zongzheng}\ \bibnamefont {Du}}, \bibinfo {author}
  {\bibfnamefont {Chuang-Han}\ \bibnamefont {Hsu}},  \emph {et~al.},\
  }\bibfield  {title} {\enquote {\bibinfo {title} {Observation of the nonlinear
  {Hall} effect under time reversal symmetric conditions},}\ }\href@noop {}
  {\bibfield  {journal} {\bibinfo  {journal} {arXiv preprint arXiv:1809.09279}\
  } (\bibinfo {year} {2018})}\BibitemShut {NoStop}%
\bibitem [{\citenamefont {Kang}\ \emph {et~al.}(2018)\citenamefont {Kang},
  \citenamefont {Li}, \citenamefont {Sohn}, \citenamefont {Shan},\ and\
  \citenamefont {Mak}}]{kang2018observation}%
  \BibitemOpen
  \bibfield  {author} {\bibinfo {author} {\bibfnamefont {Kaifei}\ \bibnamefont
  {Kang}}, \bibinfo {author} {\bibfnamefont {Tingxin}\ \bibnamefont {Li}},
  \bibinfo {author} {\bibfnamefont {Egon}\ \bibnamefont {Sohn}}, \bibinfo
  {author} {\bibfnamefont {Jie}\ \bibnamefont {Shan}}, \ and\ \bibinfo {author}
  {\bibfnamefont {Kin~Fai}\ \bibnamefont {Mak}},\ }\bibfield  {title} {\enquote
  {\bibinfo {title} {Observation of the nonlinear anomalous {Hall} effect in
  {2D} {WTe$_2$}},}\ }\href@noop {} {\bibfield  {journal} {\bibinfo  {journal}
  {arXiv preprint arXiv:1809.08744}\ } (\bibinfo {year} {2018})}\BibitemShut
  {NoStop}%
\bibitem [{\citenamefont {Xu}\ \emph {et~al.}(2018)\citenamefont {Xu},
  \citenamefont {Ma}, \citenamefont {Shen}, \citenamefont {Fatemi},
  \citenamefont {Wu}, \citenamefont {Chang}, \citenamefont {Chang},
  \citenamefont {Valdivia}, \citenamefont {Chan}, \citenamefont {Gibson} \emph
  {et~al.}}]{xu2018electrically}%
  \BibitemOpen
  \bibfield  {author} {\bibinfo {author} {\bibfnamefont {Su-Yang}\ \bibnamefont
  {Xu}}, \bibinfo {author} {\bibfnamefont {Qiong}\ \bibnamefont {Ma}}, \bibinfo
  {author} {\bibfnamefont {Huitao}\ \bibnamefont {Shen}}, \bibinfo {author}
  {\bibfnamefont {Valla}\ \bibnamefont {Fatemi}}, \bibinfo {author}
  {\bibfnamefont {Sanfeng}\ \bibnamefont {Wu}}, \bibinfo {author}
  {\bibfnamefont {Tay-Rong}\ \bibnamefont {Chang}}, \bibinfo {author}
  {\bibfnamefont {Guoqing}\ \bibnamefont {Chang}}, \bibinfo {author}
  {\bibfnamefont {Andr{\'e}s M~Mier}\ \bibnamefont {Valdivia}}, \bibinfo
  {author} {\bibfnamefont {Ching-Kit}\ \bibnamefont {Chan}}, \bibinfo {author}
  {\bibfnamefont {Quinn~D}\ \bibnamefont {Gibson}},  \emph {et~al.},\
  }\bibfield  {title} {\enquote {\bibinfo {title} {Electrically switchable
  {Berry} curvature dipole in the monolayer topological insulator {WTe$_2$}},}\
  }\href@noop {} {\bibfield  {journal} {\bibinfo  {journal} {Nature Physics}\
  }\textbf {\bibinfo {volume} {14}},\ \bibinfo {pages} {900} (\bibinfo {year}
  {2018})}\BibitemShut {NoStop}%
\bibitem [{\citenamefont {You}\ \emph {et~al.}(2018)\citenamefont {You},
  \citenamefont {Fang}, \citenamefont {Xu}, \citenamefont {Kaxiras},\ and\
  \citenamefont {Low}}]{PhysRevB.98.121109}%
  \BibitemOpen
  \bibfield  {author} {\bibinfo {author} {\bibfnamefont {Jhih-Shih}\
  \bibnamefont {You}}, \bibinfo {author} {\bibfnamefont {Shiang}\ \bibnamefont
  {Fang}}, \bibinfo {author} {\bibfnamefont {Su-Yang}\ \bibnamefont {Xu}},
  \bibinfo {author} {\bibfnamefont {Efthimios}\ \bibnamefont {Kaxiras}}, \ and\
  \bibinfo {author} {\bibfnamefont {Tony}\ \bibnamefont {Low}},\ }\bibfield
  {title} {\enquote {\bibinfo {title} {Berry curvature dipole current in the
  transition metal dichalcogenides family},}\ }\href {\doibase
  10.1103/PhysRevB.98.121109} {\bibfield  {journal} {\bibinfo  {journal} {Phys.
  Rev. B}\ }\textbf {\bibinfo {volume} {98}},\ \bibinfo {pages} {121109}
  (\bibinfo {year} {2018})}\BibitemShut {NoStop}%
\bibitem [{\citenamefont {Zhang}\ \emph
  {et~al.}(2018{\natexlab{a}})\citenamefont {Zhang}, \citenamefont {van~den
  Brink}, \citenamefont {Felser},\ and\ \citenamefont
  {Yan}}]{2053-1583-5-4-044001}%
  \BibitemOpen
  \bibfield  {author} {\bibinfo {author} {\bibfnamefont {Yang}\ \bibnamefont
  {Zhang}}, \bibinfo {author} {\bibfnamefont {Jeroen}\ \bibnamefont {van~den
  Brink}}, \bibinfo {author} {\bibfnamefont {Claudia}\ \bibnamefont {Felser}},
  \ and\ \bibinfo {author} {\bibfnamefont {Binghai}\ \bibnamefont {Yan}},\
  }\bibfield  {title} {\enquote {\bibinfo {title} {Electrically tuneable
  nonlinear anomalous {Hall} effect in two-dimensional transition-metal
  dichalcogenides {WTe$_2$} and {MoTe$_2$}},}\ }\href
  {http://stacks.iop.org/2053-1583/5/i=4/a=044001} {\bibfield  {journal}
  {\bibinfo  {journal} {2D Materials}\ }\textbf {\bibinfo {volume} {5}},\
  \bibinfo {pages} {044001} (\bibinfo {year} {2018}{\natexlab{a}})}\BibitemShut
  {NoStop}%
\bibitem [{\citenamefont {Shi}\ and\ \citenamefont
  {Song}(2018)}]{shi2018berry}%
  \BibitemOpen
  \bibfield  {author} {\bibinfo {author} {\bibfnamefont {Li-kun}\ \bibnamefont
  {Shi}}\ and\ \bibinfo {author} {\bibfnamefont {Justin~CW}\ \bibnamefont
  {Song}},\ }\bibfield  {title} {\enquote {\bibinfo {title} {Berry curvature
  switch and magneto-electric effect in {WTe$_2$} monolayer},}\ }\href@noop {}
  {\bibfield  {journal} {\bibinfo  {journal} {arXiv preprint arXiv:1805.00939}\
  } (\bibinfo {year} {2018})}\BibitemShut {NoStop}%
\bibitem [{Note1()}]{Note1}%
  \BibitemOpen
  \bibinfo {note} {We used \protect \textsc {wien2k} \cite {blaha2001wien2k}
  with the GGA functional, a $24^3$ $k$--mesh and setting $R_{MT}=2.5$ and
  $K_{max}=8.5R^{-1}_{MT}$.}\BibitemShut {Stop}%
\bibitem [{Note2()}]{Note2}%
  \BibitemOpen
  \bibinfo {note} {For each pressure we built maximally localized Wannier
  functions for $p$ orbitals of Bi, Te and I \cite
  {KUNES20101888,MOSTOFI20142309} and computed $\Omega _{n,d}$ by Wannier
  interpolation \cite {PhysRevB.74.195118} and $D_{bd}$ as $\DOTSI \intop
  \ilimits@ \protect \frac {d^3k}{(2\pi )^3} \DOTSB \sum@ \slimits@ _n \protect
  \frac {\partial \Omega _{n,d}}{\partial k_n} f_0(E_{n\protect \mathbf {k}})$
  using a $1000\times 1000\times 300$ mesh in the Brillouin-zone region that
  contains the conducting states in the range of doping analyzed. We also
  performed calculations with \protect \textsc {fplo} \cite {koepernik1999full}
  based on which we found the same evolution of $\Omega _{n,d}$ with $P$ \cite
  {Note3}.}\BibitemShut {Stop}%
\bibitem [{Note3()}]{Note3}%
  \BibitemOpen
  \bibinfo {note} {See Supplementary Material, which includes Refs. \cite
  {PhysRevB.64.153102,PhysRevB.84.041202,PhysRevB.86.085204,
  PhysRevB.87.205103,PhysRevB.94.165203}, for: $(i)$ Berry-dipole dependence on
  $\mu $ $(ii)$ discussion of DFT calculations and $(iii)$ formulae associated
  with the model.}\BibitemShut {Stop}%
\bibitem [{\citenamefont {Zhang}\ \emph
  {et~al.}(2018{\natexlab{b}})\citenamefont {Zhang}, \citenamefont {Sun},\ and\
  \citenamefont {Yan}}]{Zhang2018}%
  \BibitemOpen
  \bibfield  {author} {\bibinfo {author} {\bibfnamefont {Yang}\ \bibnamefont
  {Zhang}}, \bibinfo {author} {\bibfnamefont {Yan}\ \bibnamefont {Sun}}, \ and\
  \bibinfo {author} {\bibfnamefont {Binghai}\ \bibnamefont {Yan}},\ }\bibfield
  {title} {\enquote {\bibinfo {title} {Berry curvature dipole in {Weyl}
  semimetal materials: An ab initio study},}\ }\href@noop {} {\bibfield
  {journal} {\bibinfo  {journal} {Physical Review B}\ }\textbf {\bibinfo
  {volume} {97}},\ \bibinfo {pages} {041101} (\bibinfo {year}
  {2018}{\natexlab{b}})}\BibitemShut {NoStop}%
\bibitem [{\citenamefont {Tsirkin}\ \emph {et~al.}(2018)\citenamefont
  {Tsirkin}, \citenamefont {Puente},\ and\ \citenamefont
  {Souza}}]{PhysRevB.97.035158}%
  \BibitemOpen
  \bibfield  {author} {\bibinfo {author} {\bibfnamefont {Stepan~S.}\
  \bibnamefont {Tsirkin}}, \bibinfo {author} {\bibfnamefont {Pablo~Aguado}\
  \bibnamefont {Puente}}, \ and\ \bibinfo {author} {\bibfnamefont {Ivo}\
  \bibnamefont {Souza}},\ }\bibfield  {title} {\enquote {\bibinfo {title}
  {Gyrotropic effects in trigonal tellurium studied from first principles},}\
  }\href {\doibase 10.1103/PhysRevB.97.035158} {\bibfield  {journal} {\bibinfo
  {journal} {Phys. Rev. B}\ }\textbf {\bibinfo {volume} {97}},\ \bibinfo
  {pages} {035158} (\bibinfo {year} {2018})}\BibitemShut {NoStop}%
\bibitem [{Note4()}]{Note4}%
  \BibitemOpen
  \bibinfo {note} {Weyl points pairs related by $C_3$ naturally contribute
  equally to $d_z$. Those related by time reversal also contribute equally
  because $\protect \bf {\Omega }$ is odd under time reversal but its curl is
  even.}\BibitemShut {Stop}%
\bibitem [{Note5()}]{Note5}%
  \BibitemOpen
  \bibinfo {note} {Further details of the Weyl semimetal materials will appear
  in a forthcoming publication}\BibitemShut {NoStop}%
\bibitem [{Note6()}]{Note6}%
  \BibitemOpen
  \bibinfo {note} {This follows from Eq. \ref {eq:dip_wsm} since within the
  Weyl semimetal phase $n \propto \mu ^3$. In terms of $\mu $, the Berry-dipole
  in this regime reads $d_z\approx -\protect \frac { 4\pi \mu u_x}{3 k_0 v_y^2}
  \left (1-\protect \frac {v_y^2}{v_x^2}\right )$ and vanishes when $\mu \to
  0$.}\BibitemShut {Stop}%
\bibitem [{\citenamefont {Singh}\ \emph {et~al.}(2016)\citenamefont {Singh},
  \citenamefont {Garcia-Castro}, \citenamefont {Valencia-Jaime}, \citenamefont
  {Mu\~noz},\ and\ \citenamefont {Romero}}]{PhysRevB.94.161116}%
  \BibitemOpen
  \bibfield  {author} {\bibinfo {author} {\bibfnamefont {Sobhit}\ \bibnamefont
  {Singh}}, \bibinfo {author} {\bibfnamefont {A.~C.}\ \bibnamefont
  {Garcia-Castro}}, \bibinfo {author} {\bibfnamefont {Irais}\ \bibnamefont
  {Valencia-Jaime}}, \bibinfo {author} {\bibfnamefont {Francisco}\ \bibnamefont
  {Mu\~noz}}, \ and\ \bibinfo {author} {\bibfnamefont {Aldo~H.}\ \bibnamefont
  {Romero}},\ }\bibfield  {title} {\enquote {\bibinfo {title} {Prediction and
  control of spin polarization in a {Weyl} semimetallic phase of {BiSb}},}\
  }\href {\doibase 10.1103/PhysRevB.94.161116} {\bibfield  {journal} {\bibinfo
  {journal} {Phys. Rev. B}\ }\textbf {\bibinfo {volume} {94}},\ \bibinfo
  {pages} {161116} (\bibinfo {year} {2016})}\BibitemShut {NoStop}%
\bibitem [{\citenamefont {Kune\ifmmode~\check{s}\else \v{s}\fi{}}\ \emph
  {et~al.}(2001)\citenamefont {Kune\ifmmode~\check{s}\else \v{s}\fi{}},
  \citenamefont {Nov\'ak}, \citenamefont {Schmid}, \citenamefont {Blaha},\ and\
  \citenamefont {Schwarz}}]{PhysRevB.64.153102}%
  \BibitemOpen
  \bibfield  {author} {\bibinfo {author} {\bibfnamefont {J.}~\bibnamefont
  {Kune\ifmmode~\check{s}\else \v{s}\fi{}}}, \bibinfo {author} {\bibfnamefont
  {P.}~\bibnamefont {Nov\'ak}}, \bibinfo {author} {\bibfnamefont
  {R.}~\bibnamefont {Schmid}}, \bibinfo {author} {\bibfnamefont
  {P.}~\bibnamefont {Blaha}}, \ and\ \bibinfo {author} {\bibfnamefont
  {K.}~\bibnamefont {Schwarz}},\ }\bibfield  {title} {\enquote {\bibinfo
  {title} {Electronic structure of fcc {Th}: Spin-orbit calculation with
  {${6p}_{1/2}$} local orbital extension},}\ }\href {\doibase
  10.1103/PhysRevB.64.153102} {\bibfield  {journal} {\bibinfo  {journal} {Phys.
  Rev. B}\ }\textbf {\bibinfo {volume} {64}},\ \bibinfo {pages} {153102}
  (\bibinfo {year} {2001})}\BibitemShut {NoStop}%
\bibitem [{\citenamefont {Bahramy}\ \emph {et~al.}(2011)\citenamefont
  {Bahramy}, \citenamefont {Arita},\ and\ \citenamefont
  {Nagaosa}}]{PhysRevB.84.041202}%
  \BibitemOpen
  \bibfield  {author} {\bibinfo {author} {\bibfnamefont {M.~S.}\ \bibnamefont
  {Bahramy}}, \bibinfo {author} {\bibfnamefont {R.}~\bibnamefont {Arita}}, \
  and\ \bibinfo {author} {\bibfnamefont {N.}~\bibnamefont {Nagaosa}},\
  }\bibfield  {title} {\enquote {\bibinfo {title} {Origin of giant bulk
  {Rashba} splitting: Application to {BiTeI}},}\ }\href {\doibase
  10.1103/PhysRevB.84.041202} {\bibfield  {journal} {\bibinfo  {journal} {Phys.
  Rev. B}\ }\textbf {\bibinfo {volume} {84}},\ \bibinfo {pages} {041202}
  (\bibinfo {year} {2011})}\BibitemShut {NoStop}%
\bibitem [{\citenamefont {Sakano}\ \emph {et~al.}(2012)\citenamefont {Sakano},
  \citenamefont {Miyawaki}, \citenamefont {Chainani}, \citenamefont {Takata},
  \citenamefont {Sonobe}, \citenamefont {Shimojima}, \citenamefont {Oura},
  \citenamefont {Shin}, \citenamefont {Bahramy}, \citenamefont {Arita},
  \citenamefont {Nagaosa}, \citenamefont {Murakawa}, \citenamefont {Kaneko},
  \citenamefont {Tokura},\ and\ \citenamefont {Ishizaka}}]{PhysRevB.86.085204}%
  \BibitemOpen
  \bibfield  {author} {\bibinfo {author} {\bibfnamefont {M.}~\bibnamefont
  {Sakano}}, \bibinfo {author} {\bibfnamefont {J.}~\bibnamefont {Miyawaki}},
  \bibinfo {author} {\bibfnamefont {A.}~\bibnamefont {Chainani}}, \bibinfo
  {author} {\bibfnamefont {Y.}~\bibnamefont {Takata}}, \bibinfo {author}
  {\bibfnamefont {T.}~\bibnamefont {Sonobe}}, \bibinfo {author} {\bibfnamefont
  {T.}~\bibnamefont {Shimojima}}, \bibinfo {author} {\bibfnamefont
  {M.}~\bibnamefont {Oura}}, \bibinfo {author} {\bibfnamefont {S.}~\bibnamefont
  {Shin}}, \bibinfo {author} {\bibfnamefont {M.~S.}\ \bibnamefont {Bahramy}},
  \bibinfo {author} {\bibfnamefont {R.}~\bibnamefont {Arita}}, \bibinfo
  {author} {\bibfnamefont {N.}~\bibnamefont {Nagaosa}}, \bibinfo {author}
  {\bibfnamefont {H.}~\bibnamefont {Murakawa}}, \bibinfo {author}
  {\bibfnamefont {Y.}~\bibnamefont {Kaneko}}, \bibinfo {author} {\bibfnamefont
  {Y.}~\bibnamefont {Tokura}}, \ and\ \bibinfo {author} {\bibfnamefont
  {K.}~\bibnamefont {Ishizaka}},\ }\bibfield  {title} {\enquote {\bibinfo
  {title} {Three-dimensional bulk band dispersion in polar {BiTeI} with giant
  {Rashba-type} spin splitting},}\ }\href {\doibase 10.1103/PhysRevB.86.085204}
  {\bibfield  {journal} {\bibinfo  {journal} {Phys. Rev. B}\ }\textbf {\bibinfo
  {volume} {86}},\ \bibinfo {pages} {085204} (\bibinfo {year}
  {2012})}\BibitemShut {NoStop}%
\bibitem [{\citenamefont {Rusinov}\ \emph {et~al.}(2013)\citenamefont
  {Rusinov}, \citenamefont {Nechaev}, \citenamefont {Eremeev}, \citenamefont
  {Friedrich}, \citenamefont {Bl\"ugel},\ and\ \citenamefont
  {Chulkov}}]{PhysRevB.87.205103}%
  \BibitemOpen
  \bibfield  {author} {\bibinfo {author} {\bibfnamefont {I.~P.}\ \bibnamefont
  {Rusinov}}, \bibinfo {author} {\bibfnamefont {I.~A.}\ \bibnamefont
  {Nechaev}}, \bibinfo {author} {\bibfnamefont {S.~V.}\ \bibnamefont
  {Eremeev}}, \bibinfo {author} {\bibfnamefont {C.}~\bibnamefont {Friedrich}},
  \bibinfo {author} {\bibfnamefont {S.}~\bibnamefont {Bl\"ugel}}, \ and\
  \bibinfo {author} {\bibfnamefont {E.~V.}\ \bibnamefont {Chulkov}},\
  }\bibfield  {title} {\enquote {\bibinfo {title} {Many-body effects on the
  {Rashba-type} spin splitting in bulk bismuth tellurohalides},}\ }\href
  {\doibase 10.1103/PhysRevB.87.205103} {\bibfield  {journal} {\bibinfo
  {journal} {Phys. Rev. B}\ }\textbf {\bibinfo {volume} {87}},\ \bibinfo
  {pages} {205103} (\bibinfo {year} {2013})}\BibitemShut {NoStop}%
\bibitem [{\citenamefont {G\"uler-K\ifmmode \imath \else \i \fi{}l\ifmmode
  \imath \else \i \fi{}\ifmmode~\mbox{\c{c}}\else \c{c}\fi{}}\ and\
  \citenamefont {K\ifmmode \imath \else \i \fi{}l\ifmmode \imath \else \i
  \fi{}\ifmmode~\mbox{\c{c}}\else \c{c}\fi{}}(2016)}]{PhysRevB.94.165203}%
  \BibitemOpen
  \bibfield  {author} {\bibinfo {author} {\bibfnamefont {S\"umeyra}\
  \bibnamefont {G\"uler-K\ifmmode \imath \else \i \fi{}l\ifmmode \imath \else
  \i \fi{}\ifmmode~\mbox{\c{c}}\else \c{c}\fi{}}}\ and\ \bibinfo {author}
  {\bibfnamefont {\ifmmode \mbox{\c{C}}\else~\c{C}\fi{}etin}\ \bibnamefont
  {K\ifmmode \imath \else \i \fi{}l\ifmmode \imath \else \i
  \fi{}\ifmmode~\mbox{\c{c}}\else \c{c}\fi{}}},\ }\bibfield  {title} {\enquote
  {\bibinfo {title} {Pressure dependence of the band-gap energy in {BiTeI}},}\
  }\href {\doibase 10.1103/PhysRevB.94.165203} {\bibfield  {journal} {\bibinfo
  {journal} {Phys. Rev. B}\ }\textbf {\bibinfo {volume} {94}},\ \bibinfo
  {pages} {165203} (\bibinfo {year} {2016})}\BibitemShut {NoStop}%
\bibitem [{\citenamefont {Blaha}\ \emph {et~al.}(2001)\citenamefont {Blaha},
  \citenamefont {Schwarz}, \citenamefont {Madsen}, \citenamefont {Kvasnicka},\
  and\ \citenamefont {Luitz}}]{blaha2001wien2k}%
  \BibitemOpen
  \bibfield  {author} {\bibinfo {author} {\bibfnamefont {Peter}\ \bibnamefont
  {Blaha}}, \bibinfo {author} {\bibfnamefont {Karlheinz}\ \bibnamefont
  {Schwarz}}, \bibinfo {author} {\bibfnamefont {GKH}\ \bibnamefont {Madsen}},
  \bibinfo {author} {\bibfnamefont {Dieter}\ \bibnamefont {Kvasnicka}}, \ and\
  \bibinfo {author} {\bibfnamefont {Joachim}\ \bibnamefont {Luitz}},\
  }\bibfield  {title} {\enquote {\bibinfo {title} {wien2k},}\ }\href@noop {}
  {\bibfield  {journal} {\bibinfo  {journal} {An augmented plane wave+ local
  orbitals program for calculating crystal properties}\ } (\bibinfo {year}
  {2001})}\BibitemShut {NoStop}%
\bibitem [{\citenamefont {Kuneš}\ \emph {et~al.}(2010)\citenamefont {Kuneš},
  \citenamefont {Arita}, \citenamefont {Wissgott}, \citenamefont {Toschi},
  \citenamefont {Ikeda},\ and\ \citenamefont {Held}}]{KUNES20101888}%
  \BibitemOpen
  \bibfield  {author} {\bibinfo {author} {\bibfnamefont {Jan}\ \bibnamefont
  {Kuneš}}, \bibinfo {author} {\bibfnamefont {Ryotaro}\ \bibnamefont {Arita}},
  \bibinfo {author} {\bibfnamefont {Philipp}\ \bibnamefont {Wissgott}},
  \bibinfo {author} {\bibfnamefont {Alessandro}\ \bibnamefont {Toschi}},
  \bibinfo {author} {\bibfnamefont {Hiroaki}\ \bibnamefont {Ikeda}}, \ and\
  \bibinfo {author} {\bibfnamefont {Karsten}\ \bibnamefont {Held}},\ }\bibfield
   {title} {\enquote {\bibinfo {title} {{Wien2wannier: From linearized
  augmented plane waves to maximally localized Wannier functions}},}\ }\href
  {\doibase https://doi.org/10.1016/j.cpc.2010.08.005} {\bibfield  {journal}
  {\bibinfo  {journal} {Computer Physics Communications}\ }\textbf {\bibinfo
  {volume} {181}},\ \bibinfo {pages} {1888 -- 1895} (\bibinfo {year}
  {2010})}\BibitemShut {NoStop}%
\bibitem [{\citenamefont {Mostofi}\ \emph {et~al.}(2014)\citenamefont
  {Mostofi}, \citenamefont {Yates}, \citenamefont {Pizzi}, \citenamefont {Lee},
  \citenamefont {Souza}, \citenamefont {Vanderbilt},\ and\ \citenamefont
  {Marzari}}]{MOSTOFI20142309}%
  \BibitemOpen
  \bibfield  {author} {\bibinfo {author} {\bibfnamefont {Arash~A.}\
  \bibnamefont {Mostofi}}, \bibinfo {author} {\bibfnamefont {Jonathan~R.}\
  \bibnamefont {Yates}}, \bibinfo {author} {\bibfnamefont {Giovanni}\
  \bibnamefont {Pizzi}}, \bibinfo {author} {\bibfnamefont {Young-Su}\
  \bibnamefont {Lee}}, \bibinfo {author} {\bibfnamefont {Ivo}\ \bibnamefont
  {Souza}}, \bibinfo {author} {\bibfnamefont {David}\ \bibnamefont
  {Vanderbilt}}, \ and\ \bibinfo {author} {\bibfnamefont {Nicola}\ \bibnamefont
  {Marzari}},\ }\bibfield  {title} {\enquote {\bibinfo {title} {{An updated
  version of wannier90: A tool for obtaining maximally-localised Wannier
  functions}},}\ }\href {\doibase https://doi.org/10.1016/j.cpc.2014.05.003}
  {\bibfield  {journal} {\bibinfo  {journal} {Computer Physics Communications}\
  }\textbf {\bibinfo {volume} {185}},\ \bibinfo {pages} {2309 -- 2310}
  (\bibinfo {year} {2014})}\BibitemShut {NoStop}%
\bibitem [{\citenamefont {Wang}\ \emph {et~al.}(2006)\citenamefont {Wang},
  \citenamefont {Yates}, \citenamefont {Souza},\ and\ \citenamefont
  {Vanderbilt}}]{PhysRevB.74.195118}%
  \BibitemOpen
  \bibfield  {author} {\bibinfo {author} {\bibfnamefont {Xinjie}\ \bibnamefont
  {Wang}}, \bibinfo {author} {\bibfnamefont {Jonathan~R.}\ \bibnamefont
  {Yates}}, \bibinfo {author} {\bibfnamefont {Ivo}\ \bibnamefont {Souza}}, \
  and\ \bibinfo {author} {\bibfnamefont {David}\ \bibnamefont {Vanderbilt}},\
  }\bibfield  {title} {\enquote {\bibinfo {title} {Ab initio calculation of the
  anomalous {Hall} conductivity by {Wannier} interpolation},}\ }\href {\doibase
  10.1103/PhysRevB.74.195118} {\bibfield  {journal} {\bibinfo  {journal} {Phys.
  Rev. B}\ }\textbf {\bibinfo {volume} {74}},\ \bibinfo {pages} {195118}
  (\bibinfo {year} {2006})}\BibitemShut {NoStop}%
\bibitem [{\citenamefont {Koepernik}\ and\ \citenamefont
  {Eschrig}(1999)}]{koepernik1999full}%
  \BibitemOpen
  \bibfield  {author} {\bibinfo {author} {\bibfnamefont {Klaus}\ \bibnamefont
  {Koepernik}}\ and\ \bibinfo {author} {\bibfnamefont {Helmut}\ \bibnamefont
  {Eschrig}},\ }\bibfield  {title} {\enquote {\bibinfo {title} {Full-potential
  nonorthogonal local-orbital minimum-basis band-structure scheme},}\
  }\href@noop {} {\bibfield  {journal} {\bibinfo  {journal} {Physical Review
  B}\ }\textbf {\bibinfo {volume} {59}},\ \bibinfo {pages} {1743} (\bibinfo
  {year} {1999})}\BibitemShut {NoStop}%
\end{thebibliography}%
\end{document}